# InceptionCapsule: Inception-Resnet and CapsuleNet with self-attention for medical image Classification

***Elham Sadeghnezhad*[a], *Sajjad Salem*[b]**

[a]*e.sadeghnezhad@gmail.kntu.ac.ir*
[b]*salemsajjad@aut.ac.ir*

ARTICLE INFO

ABSTRACT

Initial weighting is significant in deep neural networks because the random selection of weights produces different outputs and increases the probability of overfitting and underfitting. On the other hand, vector-based approaches to extract vector features need rich vectors for more accurate classification. The InceptionCapsule approach is presented to alleviate these two problems. This approach uses transfer learning and the Inception-ResNet model to avoid random selection of weights, which takes initial weights from ImageNet. It also uses the output of Inception middle layers to generate rich vectors. Extracted vectors are given to a capsule network for learning, which is equipped with an attention technique. Kvasir data and BUSI with GT dataset were used to evaluate this approach. This model was able to achieve 97.62 accuracies in 5-class classification and also achieved 94.30 accuracies in 8-class classification on Kvasir. In the BUSI with GT dataset, the proposed approach achieved accuracy=98.88, Precision=95.34, and F1-score=93.74, which are acceptable results compared to other approaches in the literature.

## 1. Introduction:

Due to the success of automatic feature selection and image classification and recognition, deep learning-based approaches have been heavily used in this field. Deep learning approaches usually necessitate an extensive amount of training samples, and the availability of various large datasets has significantly advanced the field. Increasing computational capability in the fields of big data has allowed deep artificial neural networks to be used successfully for data classification and regression. In traditional neural network training, all parameters and weights are reset by error backpropagation. Increasing the number of hidden layers can cause a myriad of issues for this type of training, such as long training times, non-convergence, and the risk of local minima[1].

To overcome these issues, neural networks with randomly generated weights (NNRW) were developed. The weights between the hidden layer and the input layer are chosen randomly, while the weights between the hidden layer and the output layer are worked out analytically. According to research, the training of NNRW is significantly less complex than the training of a traditional feedforward neural network[2].

ANNs(Artificial Neural Networks)[3] have attracted considerable interest due to their powerful abilities in areas such as image processing, speech recognition, natural language processing, etc.

The success of ANN models is greatly dependent on the amount of data available, the power of the computer, and the optimization of the algorithms. Traditional ANNs make models better by changing all the weights and biases bit by bit to lower a loss function, which is the difference between what the model predicts and what actually happens. To guide parameter tuning, the effects of changes in the loss function are calculated and sent to each layer during training[1]. This technique has many significant weaknesses, including a slow rate of convergence, susceptibility to local minima, and difficulty in model selection. Sometimes, NNRW can affect local minimum problems. However, the local minimum problem and model selection uncertainty remain. Transfer learning is a good solution for these two problems.

Transfer learning is a learning problem in machine learning science whereby a previously developed model is used to perform new tasks. This technique is frequently used in deep learning, especially when it comes to issues concerning computer vision and natural language processing, where pre-trained models are used as the initial step.

Transfer learning is a type of optimization that makes rapid progress or improves performance when modeling other problems. This technique is often employed when the data is limited to constructing a new phenomenon. Consequently, we can apply deep learning models which were previously trained with large datasets and have an everyday experience with the novel occurrence under exploration and a transfer-learning model that is based on the understanding Obtained from the preceding model (in the elementary and common parts of two phenomena). Traditional data mining and machine learning algorithms can be used to make predictions by leveraging the information from labeled and unlabeled training data via statistical models. Semi-supervised classification [3, 4] [5] addresses the issue that labeled data may be insufficient to build a good classifier, in which case they use some unlabeled data for the model. Also, supervised and semi-supervised learning has been studied for many datasets. In general, it is guessed that labeled and unlabeled data have a similar distribution.

On the other hand, the areas, activities, and distributions used for training and testing are different for transfer learning. Examples of transfer learning can be found in the real world. For instance, being able to identify apples may be advantageous in recognizing pears. Similarly, playing the electronic organ aids in the teaching of piano. Transfer learning is investigated, as some observe that pre-existing knowledge can tackle new problems more expeditiously or with greater success.

The Research on transfer learning has received more attention since 1995 under various names: learning to learn, lifelong learning, knowledge transfer, inductive transfer, multi-tasking learning, knowledge integration, adversarial learning, knowledge-based inductive bias, meta-learning, and Incremental/cumulative learning [6]. Transfer learning is related to more than just deep learning issues. Instead, it can be widely used in multi-task learning or concept drift issues. However, transfer learning is prevalent in deep learning problems and provides the huge resources needed to train deep learning models. Another challenge vector deep learning models such as CapsuleNet face is the selection of rich initial vectors. These boards are usually taken from RNN networks or convolutional networks. Capsule models are able to learn more efficiently when vectors are richer. The proposed approach is presented to solve these two problems. The proposed inception capsule approach has the following learning process.

- Applying the inception model with ImageNet initial weights on the input images to extract the first vectors (Vector Extraction Phase)
- Training the capsule network on the extracted vectors in the Vector Extraction Phase, the Vector Learning Phase
- Applying Attention to learn the essential features learned in the Vector Learning Phase

The most significant achievements of this approach can be summarized as follows:

- Using transfer deep learning with initial weights of ImageNet to learn initial weights
- Using Inception to extract rich vectors for the capsule network
- Using Self-Attention to learn the best features

## 2.Background Research

In computer-aided diagnosis (CADx), several diseases can be automatically diagnosed in different parts of the body, such as the brain[7-9], breast[10], lung [9], etc. As well as diagnosing these diseases, CADx can be used to analyze endoscopic images in order to diagnose acute gastrointestinal diseases. Using CADx has various benefits, such as abbreviating the endoscopic analysis and cutting down the expense of treatment because of the early discovery of the lesion. When compared to manual examination, CADx has been shown to heighten the precision of diagnosing gastrointestinal diseases.

The utilization of deep learning (DL) methods has enabled researchers to make advances in the analysis of endoscopic images. CNN networks have been the most widely used type of network in endoscopy[11]. An algorithm based on CNN has been proposed for diagnosing ulcers and bleeding disorders of the gastrointestinal tract by the authors [12]. The two layers of VGG-16 were analyzed by their system to extract deep features. Utilizing the combination of the features, evolutionary search through PSO was used to determine the most vital features. In order to train the SVM classifier, these features were used. According to Igarashi et al. [12], the AlexNet is used to characterize various gastrointestinal diseases using a CADx framework. Initially, AlexNet used spatial features to extract the disease type and then classified them based on 14 different disease types. Based on the fully connected layer of each of these networks, features were extracted from these networks and classified separately.

A wide variety of research has been conducted on the classification, diagnosis, imaging, and endoscopy of the gastrointestinal tract (GI). In most recent research, deep learning has been used with Deep convolutional architectures such as LeNet, AlexNet, ResNet, DenseNet, and GoogleNet.

As described in[13], the authors used neural learning structures to classify images of the gastrointestinal tract. A number of networks were used for the classification process, such as the VGG16, ResNet, MobileNet, Inception-v3, and Xception networks, which were retrained for this purpose. With 8 classes in the Kvasir dataset, we can identify anatomical landmarks (pylorus, z-line, cecum), diseased states (esophagitis, ulcerative colitis, polyps), or medical procedures (dyed lifted polyps, dyed resection margins). Based on the validation data, the accuracies achieved with VGG, ResNet, MobileNet, Inception-v3, and Xception were 98.3\%, 92.3\%, 97.6\%, 90\%, and 98.2\%, respectively. Xception and VGG16 provide the most accurate results when retraining neural networks with 98% accuracy.

Transfer learning was also used as the basis of work in [14]. In this study, three networks of GoogleNet, ResNet-50, and AlexNet were used to classify Kvasir data with 5 classes: Dyed lifted polyps, normal-polyps, pylorus, normal cecum, ulcerative colitis. In this approach, the initial weights are trained using ImageNet. In the data learning process, data augmentation was used. In general, they extracted 9216 features using convolutional layers. These approaches were compared on 5000 images, 80% of which were considered as training and 20% as testing, with four evaluation criteria: Accuracy, sensitivity, specificity, and AUC. According to the AlexNet approach, the best results were obtained for Accuracy, Sensitivity, Specificity, and AUC off-validation set with values of 0.9700, 0.9680, 0.9920, and 0.9998 respectively.

Three approaches VGG16, ResNet-18, and GoogleNet, along with data augmentation, were used in [15]to classify 8 classes on Kvasir. VGG16, ResNet-18, and GoogleNet, were able to achieve the accuracy of 0.9633, 0.9027, and 0.7877, respectively. In this research, DenseNet, ResNet-18, Baseline+Inceptionv3+VGGNet, Ensemble model, linear regression approaches were considered as comparative models, and VGG16 approach achieved the highest accuracy. However, the Baseline+Inceptionv3+VGGNet approach also obtained an accuracy of 0.9611, which is almost similar to the VGG16 approach.

A much deeper approach of combining DenseNet and ResNet networks was proposed in [16]for Kvasir data. In this approach, the features extracted from 121-layer DenseNet are merged with 101-layer DenseNet in one layer called Add layer. In this approach, after combining the output of two networks, a 2d global average pooling operation is applied to it, and then the result of this output is used for classification. This technique achieved a 0.9263 accuracy rating with Kvasir v1 data.

Several studies have focused on the classification of breast cancer using ultrasound images. In this section, we discuss some relevant research works in this area. There are many deep learning-based approaches for diagnosing breast cancer using US images. These approaches are all CNN-based and are generally studies done in the last few years. Some of the prominent studies among these approaches are as follows. In this article [17], deep learning-based approaches for classifying breast US images have been extensively reviewed. Classification performance of breast US images of architectures such as AlexNet, VGG, ResNet, GoogleNet, and EfficientNet, among the most basic CNN architectures, has been compared. Then, transformer models, one of the most popular deep learning architectures these days, are examined, and performance is shown to be similar to that of CNN architectures in medical images. BUSI, the only publicly available dataset, was used in experimental studies. Experimental studies have shown that the transformer and CNN models successfully classify US images of the breast. It has been observed that the vision transformer model outperforms other models with 88.6% accuracy, 90.1% precision, 87.4% recall, and 88.7% F1 score. This study shows that deep learning architectures are successful in classifying US images and can be used in clinical experiments soon.

Ragab et al. [18]proposed the Ensemble Deep Learning Effective Clinical Decision Support System for diagnosing breast cancer. This system offers a way to assist radiologists and healthcare professionals using US images. The proposed method selects three popular deep learning models for feature extraction, and a more powerful machine learning technique is presented for breast cancer detection. In another similar study, Zhang et al.[19]presented a multitasking learning-based method for segmenting and classifying breast US images. This method uses soft and hard attention mechanisms simultaneously for the tasks. In this method, a dense CNN encoder is used in the classification process, and a decoder is also included. This decoder provides upsampling. Later, these units relate to attention-gated (AG) units with soft attention mechanisms. The proposed model follows a successful path in the classification of US images. Eroglu et al.[20] proposed a hybrid CNN-based method for diagnosing breast cancer using US images. This hybrid model has separately acquired and combined the features of AlexNet, MobileNetV2, and ResNet50 models. Thus, the proposed method has been used more successfully than single models in the diagnosis of breast cancer.

Ayana et al.[21]presented a new transfer learning-based approach that includes multiple CNN models for early detection of breast cancer from US images. He used features learned from the large natural image dataset ImageNet [22]. They then used the weights trained on the cancer cell line microscopic image dataset for transfer learning to train the breast US images. The primary purpose here is to achieve more successful results than models trained with transfer learning.

Joshi et al.[23] presented a novel approach to classifying breast US images. This approach is a new method based on deep learning for the pre-diagnosis of breast cancer. This proposed technique used data augmentation and transfer learning to increase model performance. The proposed method used two different breast US datasets. In the experimental results of this technique, the model showed accurate and fast prediction performance on the test set, and deep learning would be promising to help radiologists in clinical applications in the future.

Pourasad et al.[24]designed a method for diagnosing breast tumours on US images. The proposed approach includes six techniques to detect and segment ultrasound images. First, the features of the images were extracted using the fractal method. In contrast, the k-nearest neighbour, support vector machine, decision tree and Naive Bayes classification techniques were used to classify the images. Finally, CNN architecture was used to classify breast cancer based on direct US images. Considering the experimental results of the proposed method, deep learning techniques offer successful results.

Jabeen et al.[25]proposed a deep learning-based system for classifying breast cancer in US images. The proposed deep learning model is based on a DarkNet53 model and several deep learning techniques. The best-selected features are combined in the final stage using a new probabilistic serial approach. This method gave remarkable results on the public dataset. In conclusion, studies in the literature show that deep learning approaches can be used for early detection of breast cancer from breast US images. This approach is a new method based on deep learning for the pre-diagnosis of breast cancer. This proposed technique used data augmentation and transfer learning to increase model performance. The proposed method used two different breast US datasets. In the experimental results of this technique, the model showed accurate and fast prediction performance on the test set, and deep learning would be promising to help radiologists in clinical applications in the future.

Fang and Zeng [26]combine the convolutional neural network (CNN) with a variation-regularized model. Prior knowledge is retrieved from the noisy image. Subsequently, the edge regularization and the total variation approach are combined. Finally, the authors use the split Bregman method to clean the image. A CNN extracts the designed edge features from noisy images, while the total variation regularization method de-noises the edges.

Lee et al.[27] remove the noise at different scales by modifying the U-Net method. First, a multi-scale pyramid is created and passed to the U-Net. The residual method is applied to denoise the pyramid and produce a clean image. A performance improvement is achieved based on the coarse-to-fine segmentation.

Wu et al.[28] propose an end-to-end deep neural network. The model inputs multiple tunable noise levels and outputs a clean image. The model is trained by simultaneous bucket signals and ground truth pairs, and then the object is retrieved from experimental one-dimensional bucket signals.

Segmentation problems are solved by grouping pixels with the same structure and colour. The learning methods are categorized as supervised and unsupervised. The K-means and the FCM are examples of unsupervised methods. At the same time, the SVM, the artificial neural network (ANN), Bayesian classifier, and CNN are examples of supervised methods[29].

Vakanski, Xian, and Freer [30] combined the attention gate and the UNet. The decoding blocks use the upsampled layer, while the attention gate is used in each encoder block. Han et al. [31]propose the dual-attentive Generative adversarial network algorithm. The method includes a segmentation network and evaluation network to estimate the segmentation quality of the input. An adversary learning is arranged between the two to maximize the segmentation accuracy. Kumar et al.[32] use CNN and an improved UNet algorithm to segment the ROI. This method is a new modification of the UNet, relying on the down-sampling of the original image. A weighting pixel procedure by the majority voting produces the final output. Hiramatsu et al ,Osman and Yap [33]use a combination of deep learning and Frost filters. The improved mean shift method segments the image and is combined with the binary thresholding. The deep learning phase includes the UNet architecture and the fully connected network (FCN-AlexNet). Q. Huang et al. [34] have introduced the semantic classification of superpixels for BUS image's segmentation as follows: cropping an ROI in the base image by a selection of two diagonal points, then histogram equalization, bilateral filter and pyramid mean shift filter are applied for enhancing the image, dividing the cropped image into many superpixels by simple linear iterative clustering (SLIC), and finally the classification process by a back propagation neural network (BPNN) followed by a k-nearest neighbour (KNN) achieving the final result.

W. Gomez and W. Pereira [35] have introduced their comparative study for SS of breast tumours in ultrasound images utilizing eight well-established public convolutional neural networks (CNNs): FCN with AlexNet network, UNet, SegNet using VGG16 and VGG19, and DeepLabV3+ using ResNet18, ResNet50, MobileNet-V2, and Xception. They achieved their study's aim of selecting an efficient CNN-based segmentation model for further use in CAD systems. Their study has applied transfer learning (TL) for fine-tuning these eight CNNs to segment BUS images into two classes, normal and cancerous pixels, using more than 3000 BUS images (brought from seven ultrasound machine models) for training and validation. Based on the final performance evaluation of their study, they recommend using ResNet18 when trying to implement a fully automated end-to-end CAD system. Moreover, they have made their study's eight generated CNN models available to all researchers through a link mentioned in their paper.

K. Huang et al.[36]have introduced their study of fuzzy SS of BUS image with breast anatomy constraints by two steps: first, fuzzy FCN for good segmentation, and second, using breast anatomy-constrained conditional random fields to fine-tune the segmentation results.

Yuan Xu et al.[37]have introduced their machine learning-based work of medical BUS images' segmentation, proposing a CNNs-based fully automatic BUS images' segmentation method into four significant tissues: skin, fibroglandular tissue, mass, and fatty tissue, resulting in efficient automated segmentation providing a helpful reference to radiologists for better breast cancer characterization and breast density assessments.

K. Huang et al.[38] have introduced medical knowledge-constrained SS for BUS images, proposing an approach using information-extended images for training an FCN for SS of BUS images into three classes: cancer, mammary layer, and background, followed by applying layer structure information, locating breast cancers into the mammary layer, conducting breast cancer Segmentation by a conditional random field (CRF) producing more precise segmentation result.

Y. Lei et al. [39]have introduced their study for breast tumour segmentation in three-dimensional (3D) ABUS, proposing a developed Mask scoring region-based CNN (Mask R-CNN) consisting of five subnetworks: a backbone, a regional proposal network, a region CNN head, a mask head, and a mask score head. Their approach has been validated on 70 patients' images with ground truth manual contour, resulting in an efficient segmentation of breast cancer volume from ABUS images. S. Abbas et al.[39]have introduced an approach named BCD-WERT for breast cancer detection utilizing a Whale Optimization Algorithm (WOA) and an extremely randomized tree for enhanced selection and classification of features. When compared with eight different machine learning (ML) algorithms (Support Vector Machine (SVM), Random Forest, Kernel SVM, Decision Tree, Logistic Regression (LR), Stochastic Gradient Descent (SGD), Gaussian Naive Bayes (GNB) and k-Nearest Neighbor (KNN)), BCD-WERT has achieved an outperform- mance over all the eight.

## 3. Proposed InceptionCapsule

The components of CNNs are: Convolution layer, Pooling layer, and Density layer (fully connected). Every time we want to add a new layer before the compression layers (which are at the end of the network), we need to determine two important points: choosing between convolution and pooling operations and determining the number and size of filters from the output of the previous layer that will enter the new layer.The ideal solution is to be able to try all the available options in one layer at once. In this regard, the Google research team designed a new architecture that has a new layer called Inception. The main goal of designing the Inception module was to create several operations (pooling, convolution) with filters of different sizes ($3*3$, $5*5$, etc.) in parallel, and there is no need to choose between them. The initial module of the Inception-v1 architecture is one of the most fundamental parts of the system, having undergone further refinement since 2015.

There are Max-pooling layers in the GoogleNet network that reduce its resolution by halving. Adding $1*1$ convolutional layers to the network increases the depth of the network and this has been a key method used in Inception-ResNet[40]. It should be noted that both Inception-v2 and v3 were introduced in [41]. [42] introduced batch normalization which was used in Inception-v2 and [43] added factorization in version 3. Inception-v4 is more improved than v3 [43]. In[43], they also introduced Inception-ResNet, which we modified and used in this research work. Residual connectivity enables the proposed model to grow significantly while maintaining superior performance and not overfitting like other deep neural networks.

In our approach, we use this Inception-ResNet network to perform the feature extraction task and have used only a few parts of the original architecture from[43].

Despite the advancements in deep learning approaches, they could not create effective classifications when the model is scaled, leading to overfitting and decreased performance. To achieve a more accurate classification task, we propose a model called InceptionCapsule, based on the success of the Inception-ResNet model, a model that has been used successfully in other deep learning fields. This Inception capsule model is derived from the Inception-ResNet model, with the key components being the residual connections and the Inception module. Figure 1 shows the implementation of the Inception module using the TensorFlow Keras API. The layers of the proposed approach are as follows:

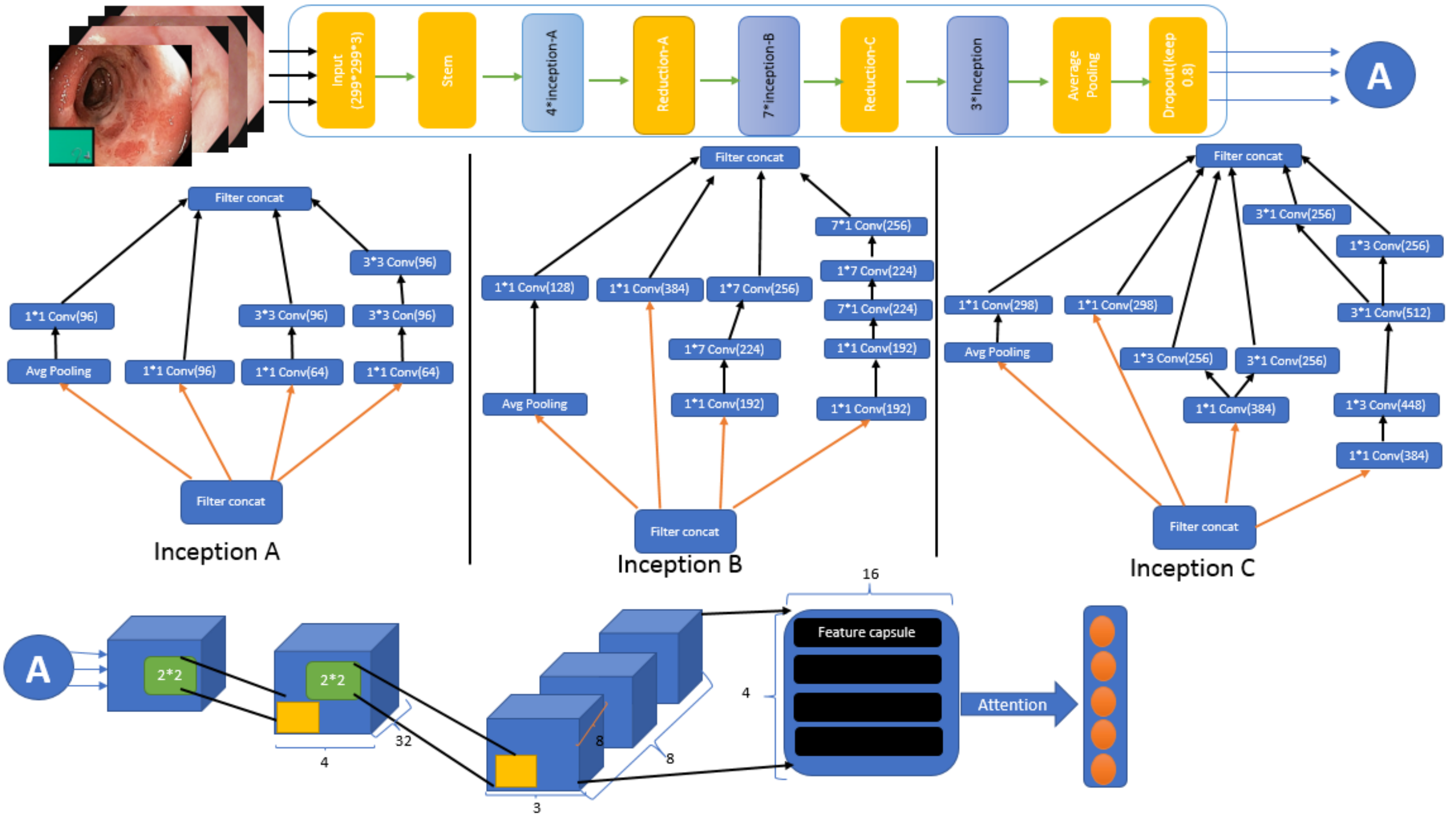

Figure 1: The proposed Inception capsule with attention.

**Inception ResNet layer:** In order to build a very deep and extensive model, basic modules are very beneficial. The Inception modules run different convolutions with different kernel sizes simultaneously, after which the results of these parallel operations are connected. The preceding layer gives the input for this block. It is connected to all four 1*1 1-dimensional convolutional layers (Conv1D) and Maxpool 1D, a 1*3 max pooling operation. This 1*1 Conv1D is an inexpensive option that reduces the dimensions of the input features while maintaining their significance. The process of removing the extra channel is considerably more economical, as demonstrated in [41] and [43]. This 1*1 Conv1D is referred to as a bottleneck which diminishes the input channels to 1.

In Inception A, the input works for 3 more convolution layers (1*1 and two 1*3). Next, the output is sent to a reducing layer. After that, Inception B and Inception C are applied to it. The details of each of these blocks are given in Figure 1. In the following, an Average Pooling and a Dropout with a Drop rate of 0.80 are applied to it. At the end of this process, the Dropout output is used as a feature selection layer, and its output is given to the capsule layer in vectors.

**Capsule layer (Caps):** Inception ResNet encoded features are given to the Capsule layer. [44]defined a capsule as a group of neurons with sampling parameters represented by activity vectors, where the length of the vector represents the probability of feature presence. This network is comprised of a convolutional layer, a primary capsule layer (PC), and a class capsule layer. The first capsule layer is the initial capsule layer, which is followed by an unknown number of capsule layers until the last capsule layer, which is also called the class capsule layer. As a result of the convolution layer, the image's features are extracted and sent to the first capsule layer. Each capsule i (where $1 \leq i \leq N$) in layer $l$ has an activity vector $u_i \in R$ to encode spatial information in the form of sample parameters. The output vector $u_i$ of the lower-level capsule $i$ is fed to all capsules in the next layer $l+1$. The jth capsule in layer $l+1$ receives $u_i$ and finds its product with the corresponding weight matrix $w_{ij}$ [39]. The resulting vector $\hat{u}_{j|i}$ is capsule $i$ in the transformation of level $l$ of the entity represented by capsule $j$ at level $l+1$. The prediction vector of a PC, $\hat{u}_{i|j}$shows how much the initial capsule $i$ contributes to the class capsule $j$.

This layer includes a set of capsules. The Caps convert the scalar features extracted by the Inception ResNet layer into vector-valued capsules to capture the input sequence features. If InceptionResNet output is $h_i$, and $w$ is a weighted matrix, then $\hat{t}_{i|j}$ which represents the predictor vector, is obtained from the following equation:

$$\hat{t}_{i|j} = w_{ij} h_i$$

The set of inputs to a capsule $Z_j$ is a weighting set of all prediction vectors $\hat{t}_{i|j}$, which is computed according to the following equation:

$$Z_i = \sum_{i=1}^{N} c_{ij}.\hat{t}_{i|j}$$

Where $c_{ij}$ is the coupling coefficient, which is repeatedly adjusted by Dynamic Routing algorithm[44]. The "squash" is used as a nonlinear function for mapping the values of $Z_j$ vectors to [0-1]. This function is applied to $Z_j$ according to the following equation:

$$v_j = \frac{\left|\left|Z_j\right|\right|^2 Z_j}{1 + \left|\left|Z_j\right|\right|^2 ||Z_j||}$$

The output of a capsule is a vector that can be sent to one of the higher-level capsules that have been chosen. In the proposed architecture, Dynamic Routing was used as the routing mechanism.

**Self-attention layer**: An attention mechanism for encoder-decoder models was introduced to improve machine translation performance. A weighted combination of all encoded input vectors, assigning the highest weights to the most relevant vectors, was the basis for the attention mechanism, which allowed the decoder to take advantage of the most relevant parts of the input sequence flexibly. The aim of image mining is to pay attention to the most significant parts of the image. The mapping of attention is:

$$Attention(q, K, V) = \sum_{i=1}^{K} softmatch_a(q, k)_i . v_i$$

Where $q \in Q$ a query, $Q \subseteq R_q^d$ the query-space, $K \subseteq R_k^d$ the key-space and $K = \{k_1, \ldots, k_N\} \subseteq K$, $V \subseteq R_v^d$ the value-space and $V = \{v_1, \ldots, v_N\} \subseteq V$, and $softmatch_a(q, k)_i$ is a probability distribution over the elements of $K$ defined as:

$$softmatch_a(q, k)_i := \frac{\exp\left(a(q, k_i)\right)}{\sum_{j=1}^{N} \exp\left(a(q, k_i)\right)} = softmax_j(a(q, k_i))$$

Moreover, when $K = V = Q$ self-attention can be defined as:

$$Q \longmapsto SelfAttention(Q) := Attention(Q, Q, Q)$$

**Classification Layer**: The flattened outputs of the self-attention layer, represented by $F$, are given to a fully connected layer of $n$ neurons (n=number of class).

$$P = W_{dense} * F$$

The output of $P$ should be such that it represents the probability of each of the $n$ class. For this purpose, we use the Softmax function, which is calculated for each $f_i \subseteq F$ as follows:

$$P_i = \frac{e^{-f_i}}{\sum_{f_{j \in f}} e^{-f_i}}$$

Where $\boldsymbol{l}$ represents the loss function and is a measure of the complexity of the model and can help prevent the model from over-fitting.

## 4. results

The proposed method was trained on a computer with a core i7-12700k CPU (2.7 GHZ), 64 GB DD RAM and NVIDIA GeForce GTX 1090 Ti. Also, the software configurations of the proposed approach are given in the Table 1:

Table 1: Software Configuration.

| Application | Version |
|---|---|
| Operation System | Linux Ubuntu |
| Python | Version 3.9 |
| TensorFlow | Version 2.0 |
| Keras | Version 2.0 |

The following three criteria were used to evaluate the proposed approach:

$$Accuracy = \frac{\mathrm{TP} + TN}{\mathrm{TP} + \mathrm{FN} + \mathrm{TN} + \mathrm{FP}}$$

$$Precision = \frac{TP}{TP + FP}$$

$$Recall = \frac{TP}{TP + FN}$$

$$Sensitivity = \frac{TP}{TP + FN}$$

$$Specificity = \frac{TN}{TN + FP}$$

where True Positive (TP) is the number of correct images that have been correctly detected, False Positive (FP) is the number of false images that have been correctly detected, True Negative (TN) is the number of correct samples that have been detected as incorrect, and False Negative (FN) is the number of false samples that are wrongly recognized. Also, the proposed approach was investigated in two modes of 5 classes classification, and 8 classes classification. In the following, the results of each of these classifications will be examined.

**4.1 Kvasir 5 class classification**

In the classification of 5 classes, the proposed approach was compared with 6 previous studies. These studies are as follows:

- [45]: This approach is a 3-layer convolutional network in which a Maxpooling layer is added after each convolution and Softmax is used as a loss function in the last layer.
- [46]: This approach is also a convolutional UNet network that is trained on global features extracted from input images, and these features include: JCD, Tamura, ColorLayout, EdgeHistogram, AutoColorCorrelogram, and PHOG.
- [47]: Different transfer learning structures such as ResNet50, ResNet101, VGG19, Xception, and Inception v3 were used in this study. ResNet50 obtained the highest accuracy and was used to compare with the proposed approach.
- [48]: The structure is a combination of CNN and SVM, CNN was used for Feature Transfer and Feature Vector Generation, and in this approach, SVM was used for the feature learning process.
- [49]: This is a GAP-Resnet50 approach that uses the ResNet50 network and Global Average Pooling (GAP) for classification. In fact, in this approach, ResNet50 is responsible for feature extraction, and the end of this architecture is given to a GAP to reduce the feature space.
- [14]: This study includes three approaches: AlxNet, GoogleNet, and Resnet50. the main contribution of the present paper is the provision of methods for lower gastrointestinal diseases with modified criteria for extracting deep shape, colour, and texture features and adapting them to learning transfer techniques for fine-tuned and contoured transfers.

The InceptionCapsule approach without self-attention was able to achieve 97.37, 99.98, and 99.49 in accuracy, sensitivity, and specificity, respectively, which is 8.81, 11.12, and 12.14 improvements in accuracy, sensitivity, and specificity compared to the [45] approach, respectively (see Table 2). This approach was also able to improve 9.37, 6.89, and 12.14 compared to the approach of [46] to obtain. Compared to the approach of [47] also this approach obtained 10.27, 12.79, and 6.49 improvements in these three criteria. Considering that the specificity value for the approach of [48] unavailability InceptionCapsule without self-attention was able to get 11.47 in accuracy and 12.29 in sensitivity. In comparison, [50] 6.41, 4.73, and 25.3 improvement in obtaining the biggest difference in the specificity approach in this approach. [51] is also one of the approaches that the proposed framework has the most differences with, and the proposed approach has 18.97, 16.59, and 29.39 improvements over this approach, respectively. [52] was also among the first approaches that were proposed for this data set and its 5 classes, compared to this approach, the proposed approach obtained 12.37, 16.89, and 16.95 in the evaluation criteria, respectively. The InceptionCapsule without self-attention approach is in equal conditions with [49] was able to obtain improvements of 7.17, 9.79, and 9.19. In the three approaches proposed in [14] the proposed approach was able to achieve 0.37, 0.67, and 2.37 improvement in accuracy compared to the modified (AlexNet), modified (GoogleNet), and modified (ResNet 50) approaches. By maintaining this order of models, the proposed approach was able to achieve 3.09, 3.29, and 5.09 improvements in the sensitivity evaluation criterion, while this improvement for the Specificity criterion was equal to 0.29, 0.69, and 0.69.

The InceptionCapsule with self-attention approach was able to achieve 97.62, 99.85, and 99.32 in accuracy, sensitivity, and specificity, respectively, which is an improvement of 9.06, 11.08, and 11.97 respectively in the accuracy, sensitivity, and specificity compared to the approach of [45](see Table 2). This approach was also able to improve 9.62, 6.85, and 11.97 compared to the approach of [46] to obtain. Compared to the approach of [47] also this approach obtained 10.52, 12.75, and 6.32 improvements in these three criteria. Considering that the specificity value for the approach of [48] unavailability InceptionCapsule with self-attention was able to get 11.72 in accuracy and 12.25 in sensitivity. In comparison, [50] 6.66, 4.69, and 25.13 improvement in obtaining the biggest difference in the specificity approach in this approach. [51] is also among the approaches to which the proposed framework has the most differences, and the proposed approach has 19.22, 16.55, and 29.32 improvements to this approach, respectively. The proposed approach compared to [52] obtained 12.62, 16.85, and 16.78 in the evaluation criteria, respectively. The InceptionCapsule without self-attention approach is in equal conditions with [49] was able to obtain improvements of 7.42, 9.75, and 9.02. In the three approaches proposed in [14] the proposed approach was able to achieve 0.62, 0.92, and 2.62 improvement in accuracy compared to the modified (AlexNet), modified (GoogleNet), and modified (ResNet 50) approaches. By maintaining this order of models, the proposed approach was able to achieve 3.05, 3.25, and 5.05 improvements in the Sensitivity evaluation criterion, while this improvement for the Specificity criterion was equal to 0.12, 0.32, and 0.52. Compared to InceptionCapsule without self-attention, this approach has improved in accuracy, but in two evaluation metrics, this approach has performed poorly.

Table 2:Optioned result on 5-class data

| Model | Accuracy (%) | Sensitivity (%) | Specificity (%) |
|---|---|---|---|
| **[45]** | 88.56 | 88.77 | 87.35 |
| **[46]** | 88.00 | 93.00 | 82.00 |
| **[47]** | 87.10 | 87.10 | 93.00 |
| **[48]** | 85.90 | 87.60 | - |
| **[50]** | 90.96 | 95.16 | 74.19 |
| **[51]** | 78.40 | 83.30 | 70.10 |
| **[52]** | 85.00 | 83.00 | 82.54 |
| **[49]** | 90.20 | 90.10 | 90.30 |
| **Modified (AlexNet) [14]** | 97.00 | 96.80 | 99.20 |
| **Modified (GoogleNet) [14]** | 96.70 | 96.60 | 99.00 |
| **Modified (ResNet-50) [14]** | 95.00 | 94.80 | 98.80 |
| **InceptionCapsule with self-Attention** | 97.62 | 99.85 | 99.32 |
| **InceptionCapsule without self-Attention** | 97.37 | 99.89 | 99.49 |

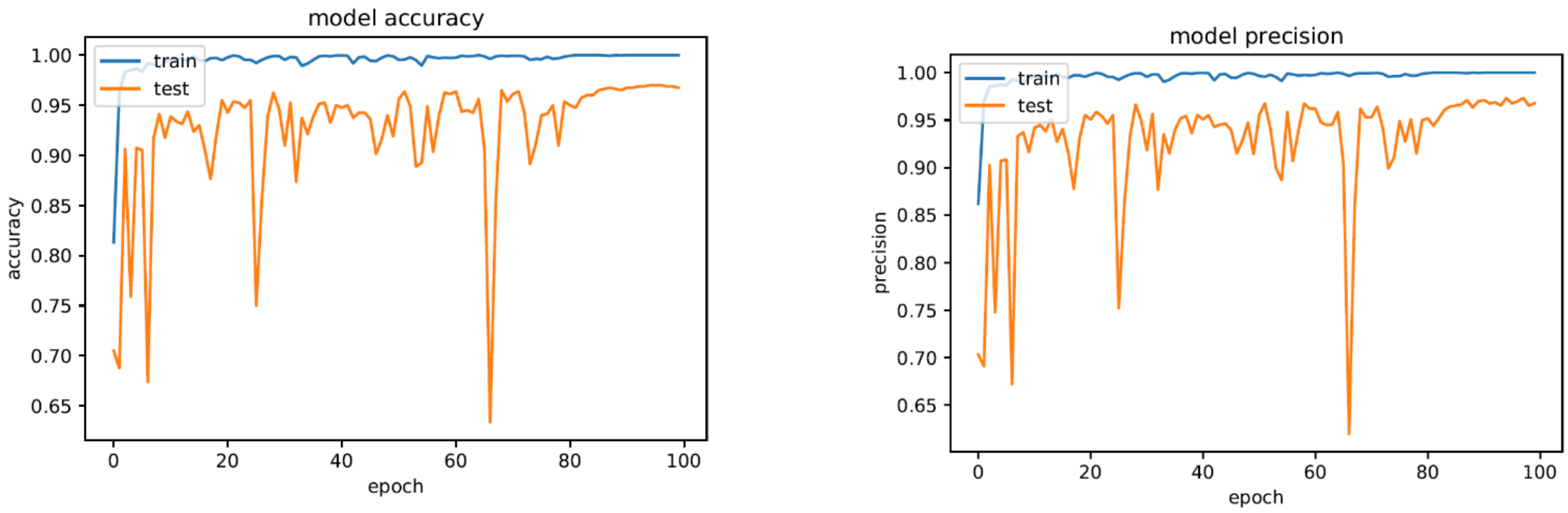

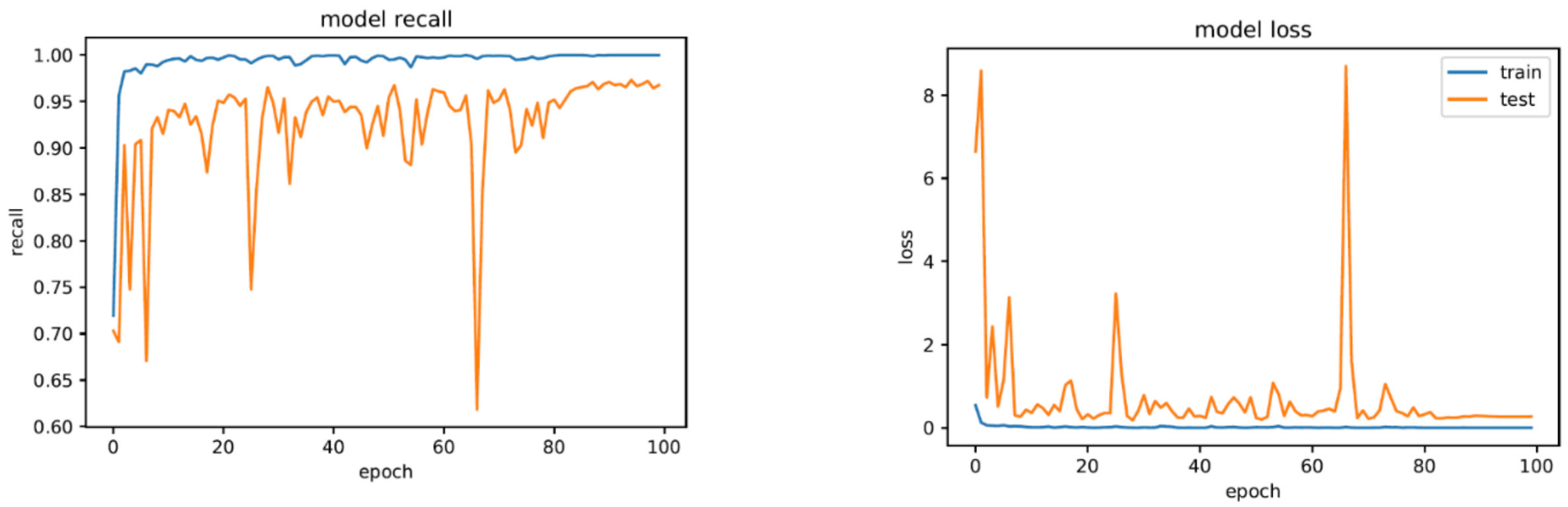

Figure 2: InceptionCapsule with self-attention plots (5 class classification)

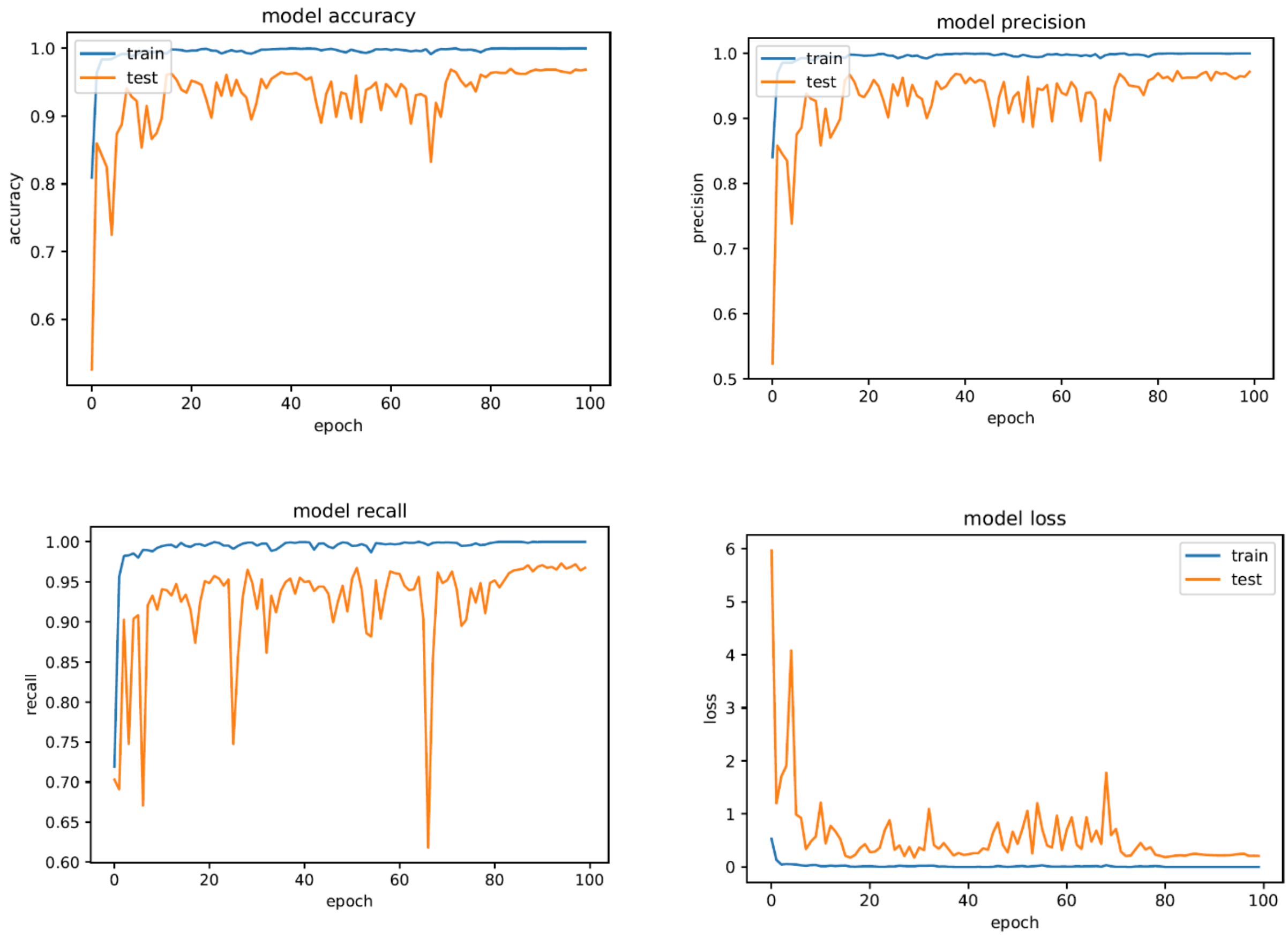

Figure 3: InceptionCapsule without self-attention plots (5 class classification)

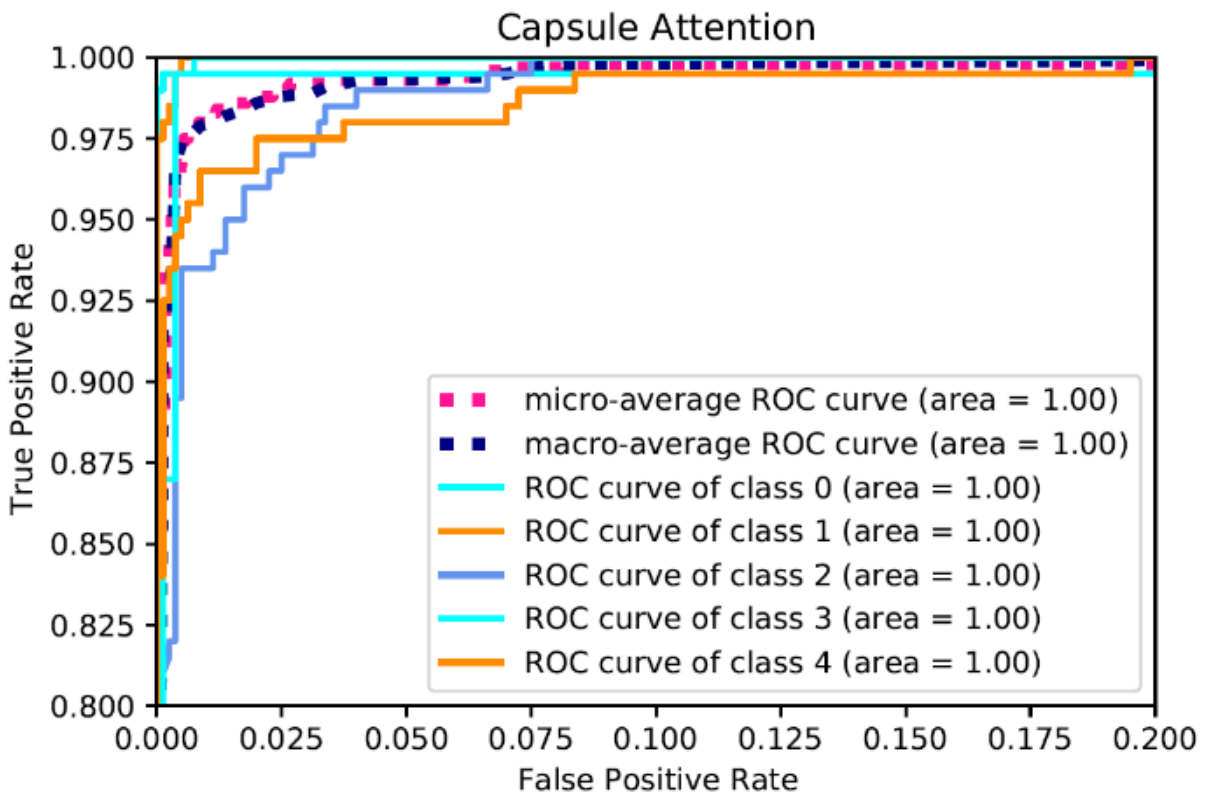

Figure 4: InceptionCapsule with self-attention

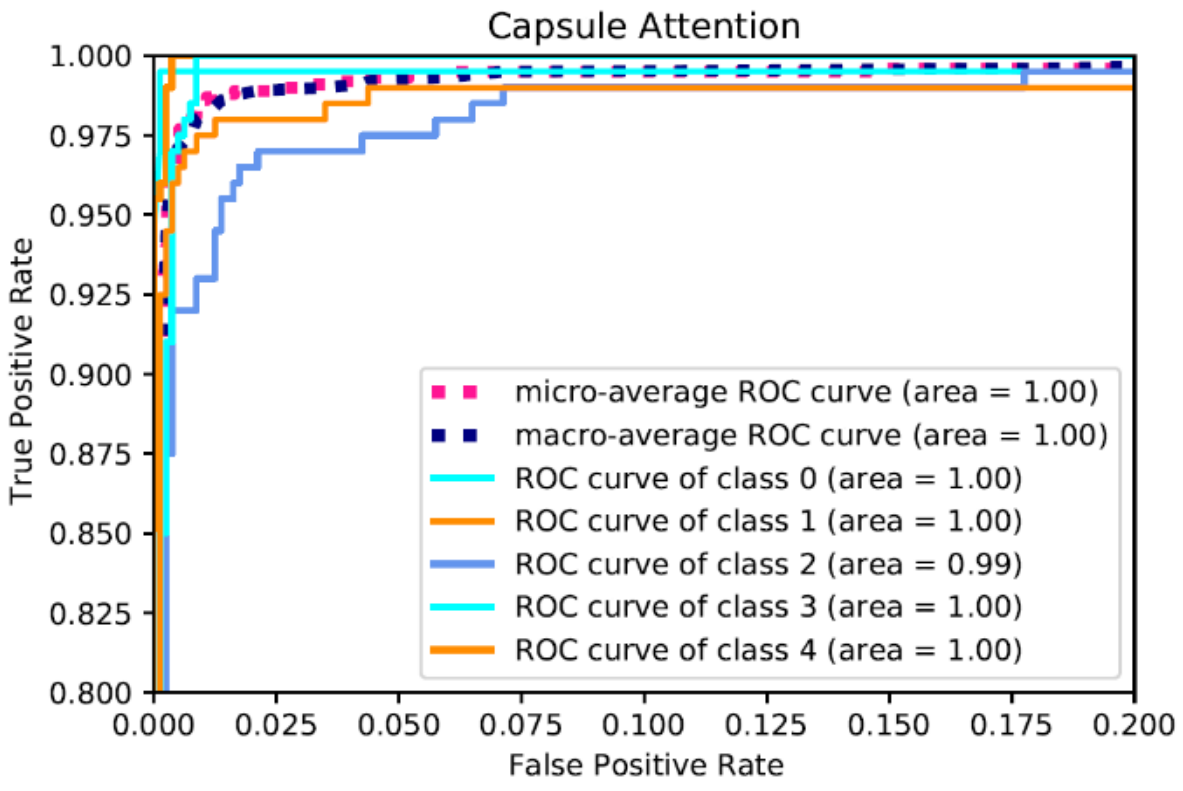

Figure 5: InceptionCapsule without self-attention

Figure 6: InceptionCapsule with self-attention and InceptionCapsule without self-attention ROC plots (5 class classification).

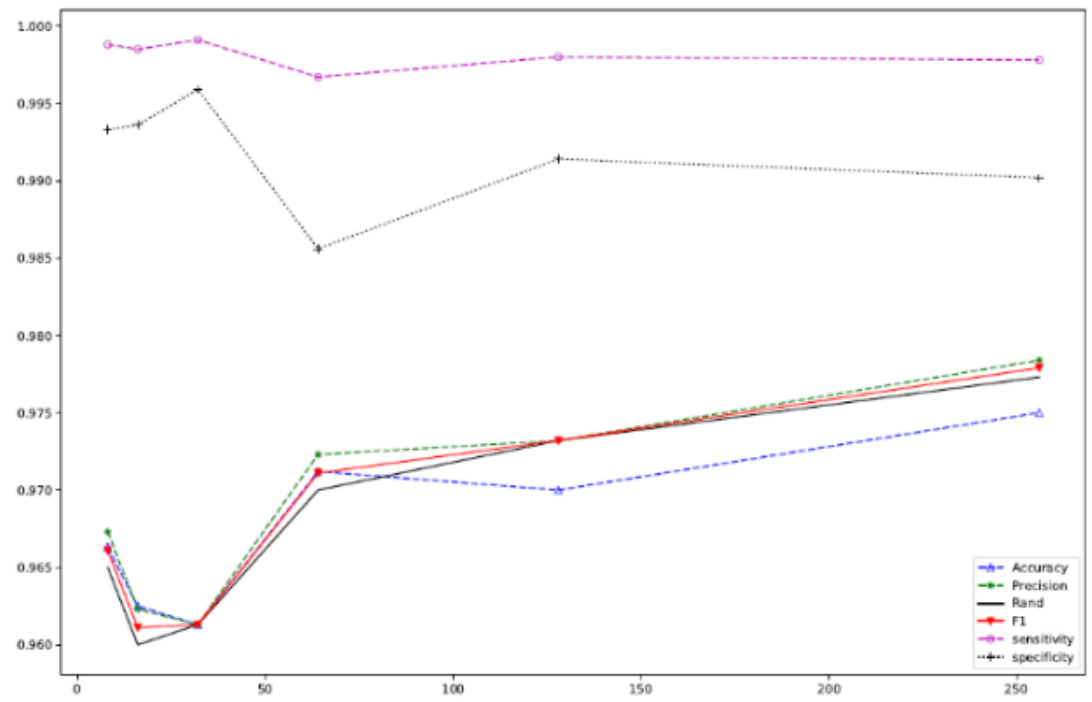

Figure 8: InceptionCapsule with self-attention

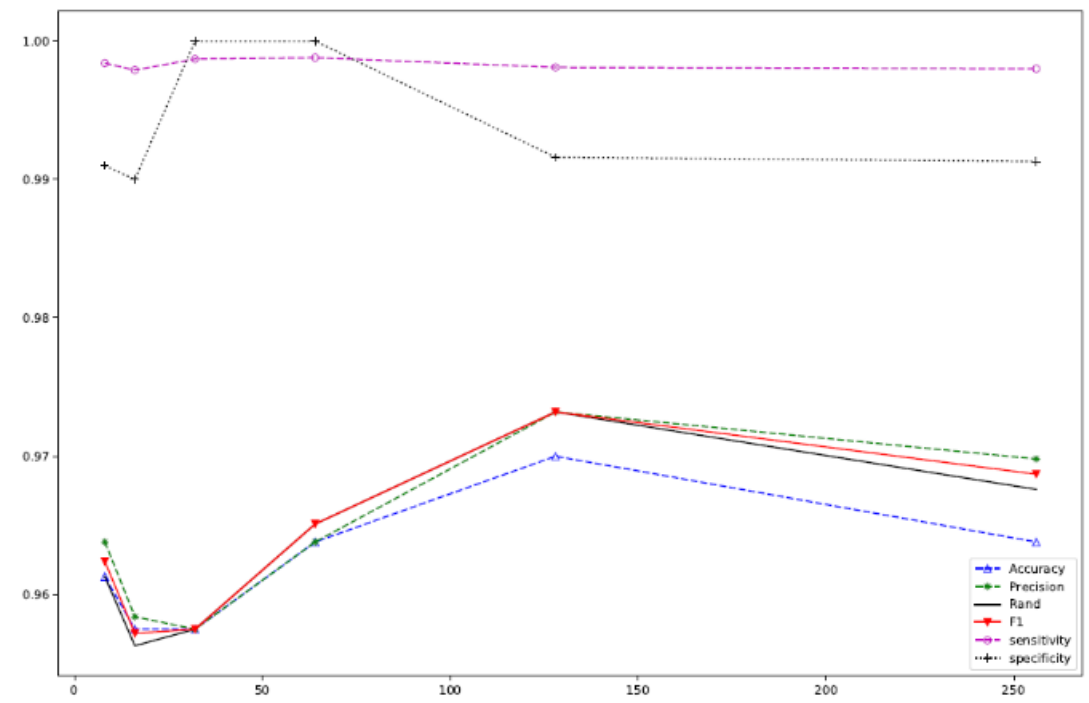

Figure 9: InceptionCapsule without self-attention

Figure 7: InceptionCapsule with self-attention and InceptionCapsule without self-attention batch size plots (5 class classification).

### 4.2 Kvasir 8 class classification

The proposed approach in the classification of 8 classes was also compared with four other approaches:

- [51]: This approach includes two basic steps of bootstrapping and fine-tuning, which used Yolo3 and RetinaNet for bootstrapping detection. In fine-tuning, this approach uses a convolutional network with transfer learning and data augmentation.
- [53]: This approach is a 3D-FCN model that is used to learn spatiotemporal feature representations. The author proposed an integrated framework with online and offline 3-D representation learning to reduce the number of false positives (FPs).
- [54]: This approach uses a ResNet18 network to extract features and SVM to learn features, this approach also uses transfer learning to learn initial weights.

In the analysis of the proposed approach of InceptionCapsule without self-attention, this approach was able to obtain 94.30, 99.87, and 99.88 respectively in accuracy, sensitivity, and specificity. Compared to the approach of [51] this approach achieved 3.44, 28.27, and 2.88 improvements in the compared evaluation criteria, respectively. This approach also compared to [55] obtained 5.36, 28.47, and 5.48 improvements in these criteria. Compared to the approach of [53] this approach obtained 3.30 and 18.87 in accuracy and sensitivity, respectively, and the specificity value of this approach is not available for comparison. [54] approach, which is one of the most recent approaches proposed for this data set, obtained results close to the InceptionCapsule without self-attention approach. InceptionCapsule without self-attention obtained only 0.92, 4.92, and 2.48 improvement compared to this approach, the difference in accuracy is very small.

In the analysis of the proposed approach of InceptionCapsule with self-attention, the difference in criteria is comparable with other approaches and in some cases it is admirable. This approach was able to obtain 95.89, 99.88, and 99.06 for accuracy, sensitivity, and specificity, respectively, in the classification of 8 classes. This approach compared to [51] 5.03, 28.28, and 2.87 to improve. Compared to [55] InceptionCapsule with self-attention obtained a great improvement, this approach obtained 6.95, 28.47, and 5.48 improvement in these three periods. Compared to [53] this approach was able to achieve 4.89 improvement in accuracy, which was less than the [54] approach, the obtained accuracy difference is 2.51.

In the plot analysis for both approaches based on the evaluation criteria of accuracy, precision, and recall as well as loss, reports were presented in two modes of 5 classes and 8 classes. In the case of 5 classes in the plots related to the InceptionCapsule with self-attention model, a sudden jump occurs in the model in epoch 66. This sudden jump is caused by the chosen values for the learning rate, which is set to 0.001. In this sudden jump, the values increase the error and decrease Precision, Recall, and Accuracy. In the case of InceptionCapsule without self-attention, this jump is relatively less. Examining the ROC diagram of the two proposed approaches shows the correctness of the proposed approaches in identifying classes with ROC curve values of 1 in most cases in both non-attention and with attention states.

In the 8-class mode, the proposed approach recorded a sudden jump, like the 5-class InceptionCapsule with self-attention mode. In this model, the examination of the ROC diagram shows that this diagram has obtained a ROC curve above 0.98 in most cases, and the lowest value of this criterion occurred in the without self-attention mode on the data of class 6.

In the graphical reports, different Batch size values for the proposed approach were presented. Different values of 8, 16, 32, 64, 128, and 256 were tested for both modes with self-attention and without self-attention. In the mode with self-attention, the highest accuracy is related to the batch size=128, and in the mode without self-attention, the highest accuracy is obtained in the batch size of 256. For further interpretation of different approaches and the proposed approach, bar graphs were used. The bar graph of the proposed approach in two modes of 5 classes and 8 classes is given in Figures 10 and 11.

Table 3:Optioned result on 8-class data.

| Model | Accuracy (%) | Sensitivity (%) | Specificity (%) |
|---|---|---|---|
| [51] | 90.86 | 71.6 | 97.0 |
| [55] | 88.94 | 71.4 | 94.4 |
| [53] | 91.0 | 71.0 | N/A |
| [54] | 93.38 | 94.95 | 97.4 |
| InceptionCapsule with self-Attention | 95.89 | 99.88 | 99.06 |
| InceptionCapsule without self-Attention | 94.30 | 99.87 | 99.88 |

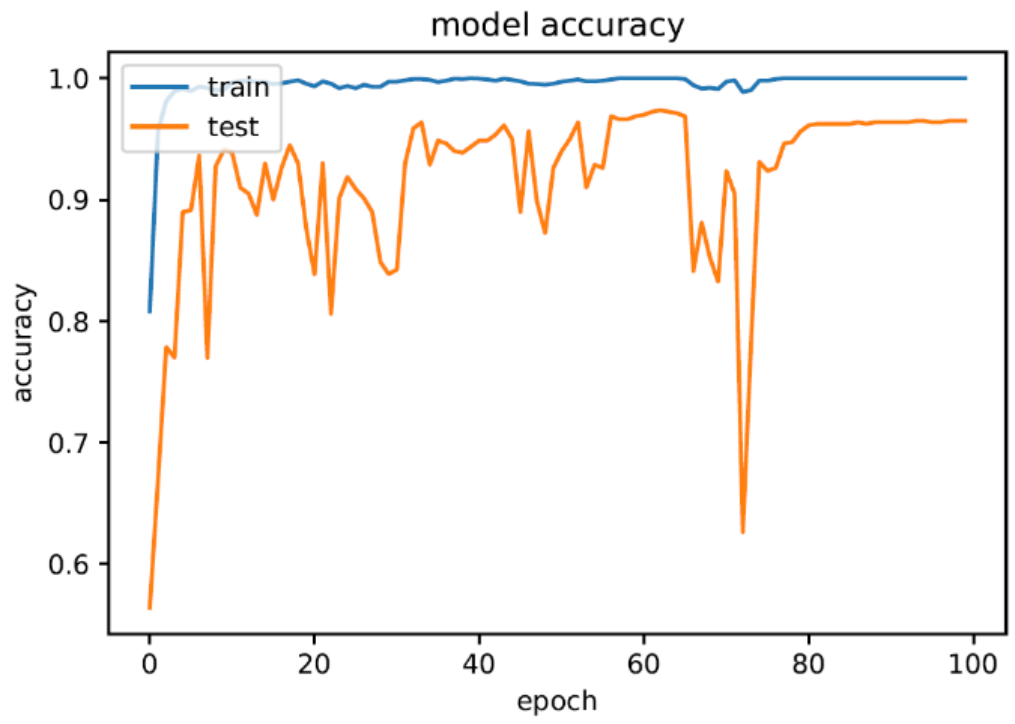

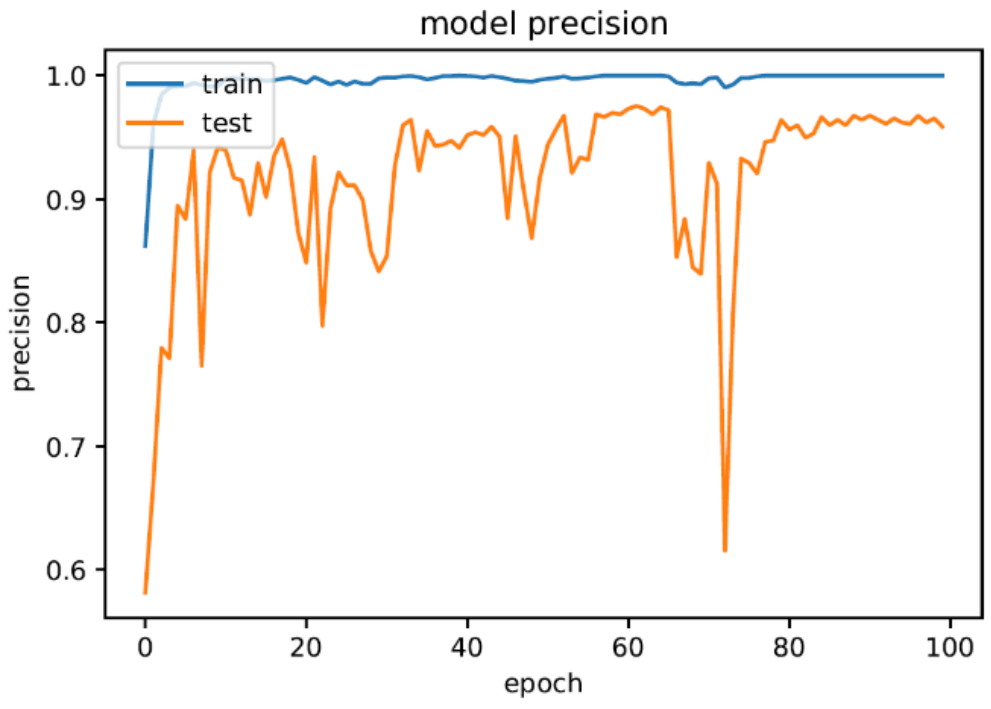

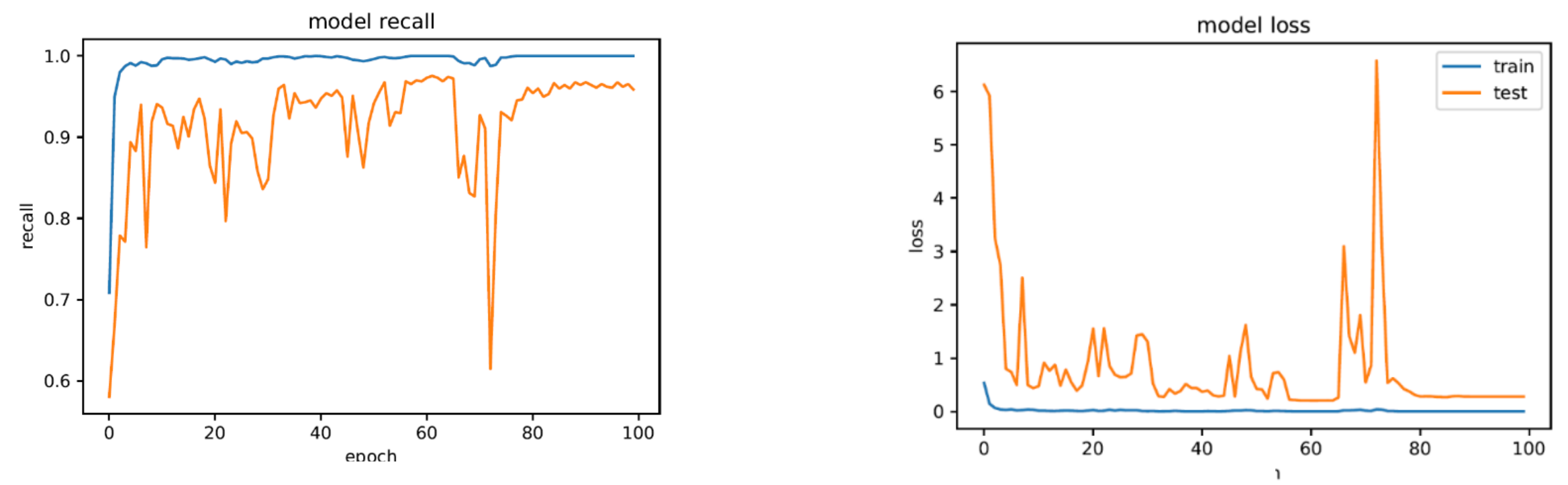

Figure 10: InceptionCapsule with self-attention plots (8 class classification)

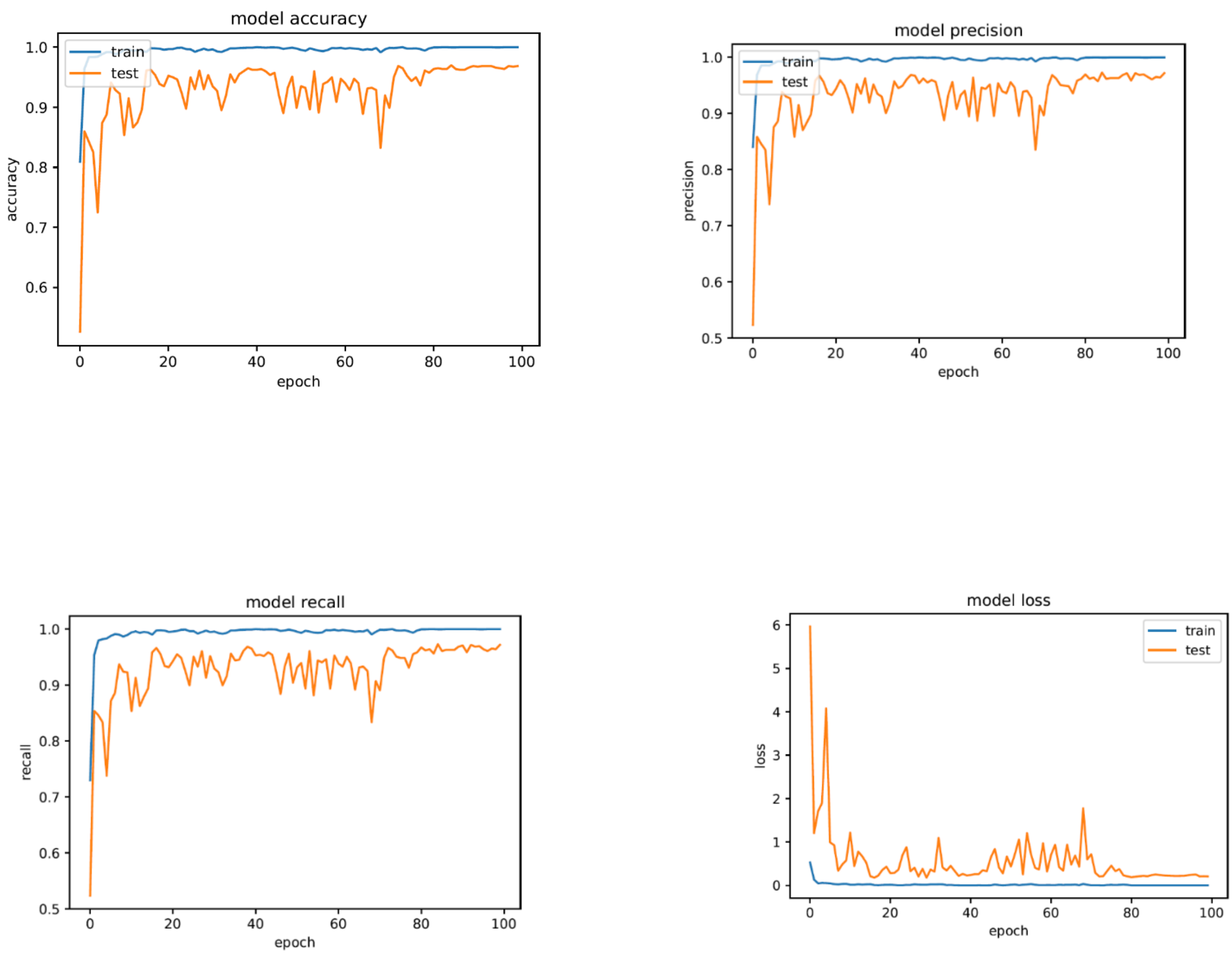

Figure 11: InceptionCapsule without self-attention plots (8 class classification)

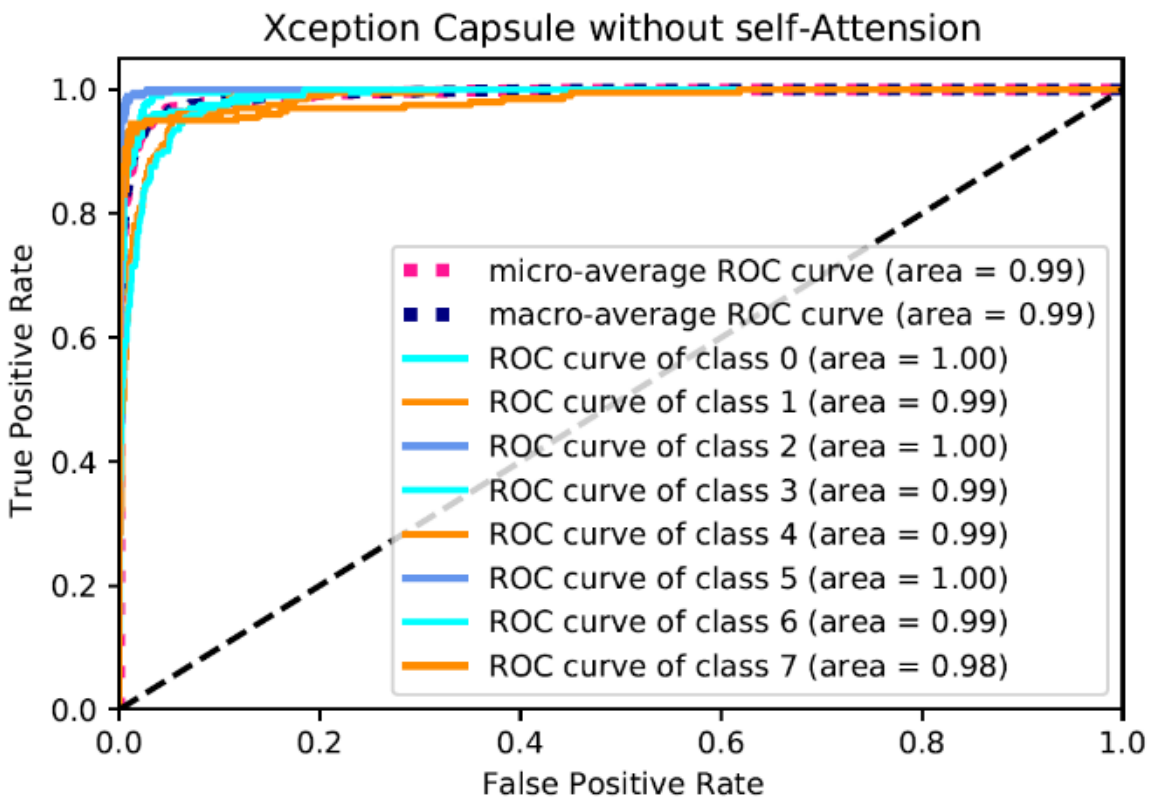

Figure 12: InceptionCapsule with self-attention

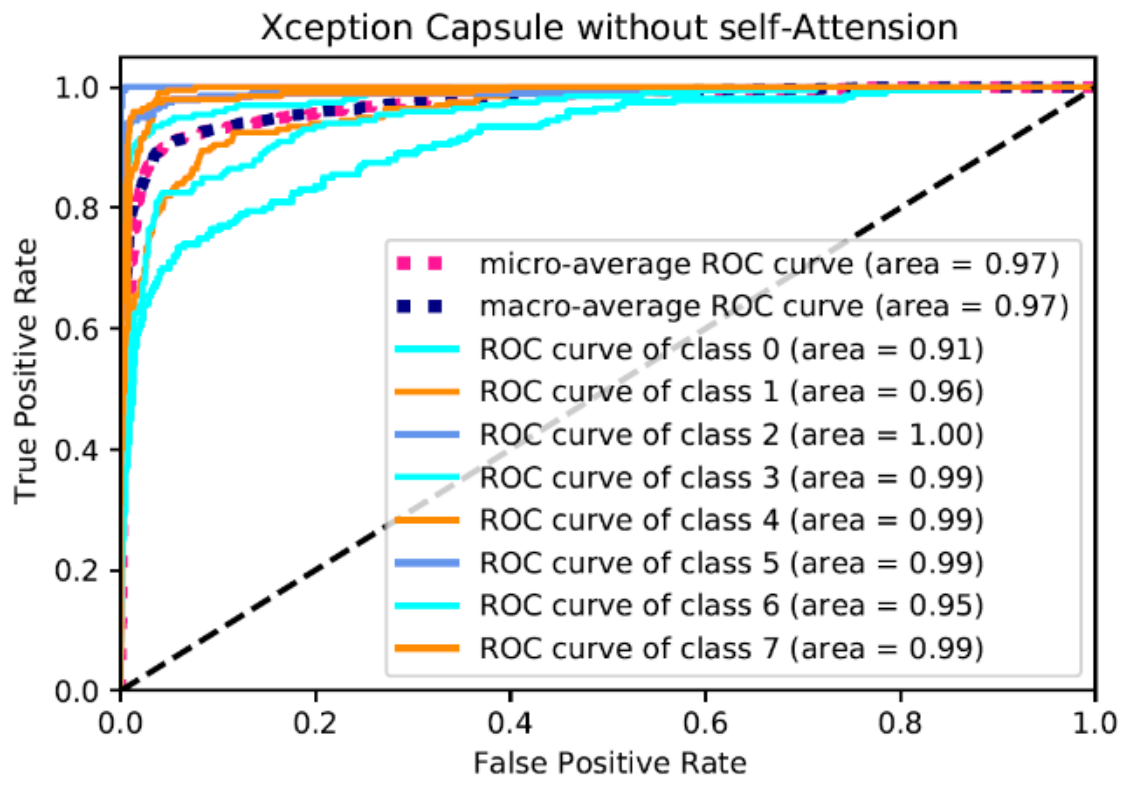

Figure 13: InceptionCapsule without self-attention

Figure 14: InceptionCapsule with self-attention and InceptionCapsule without self-attention ROC plots (8 class classification).

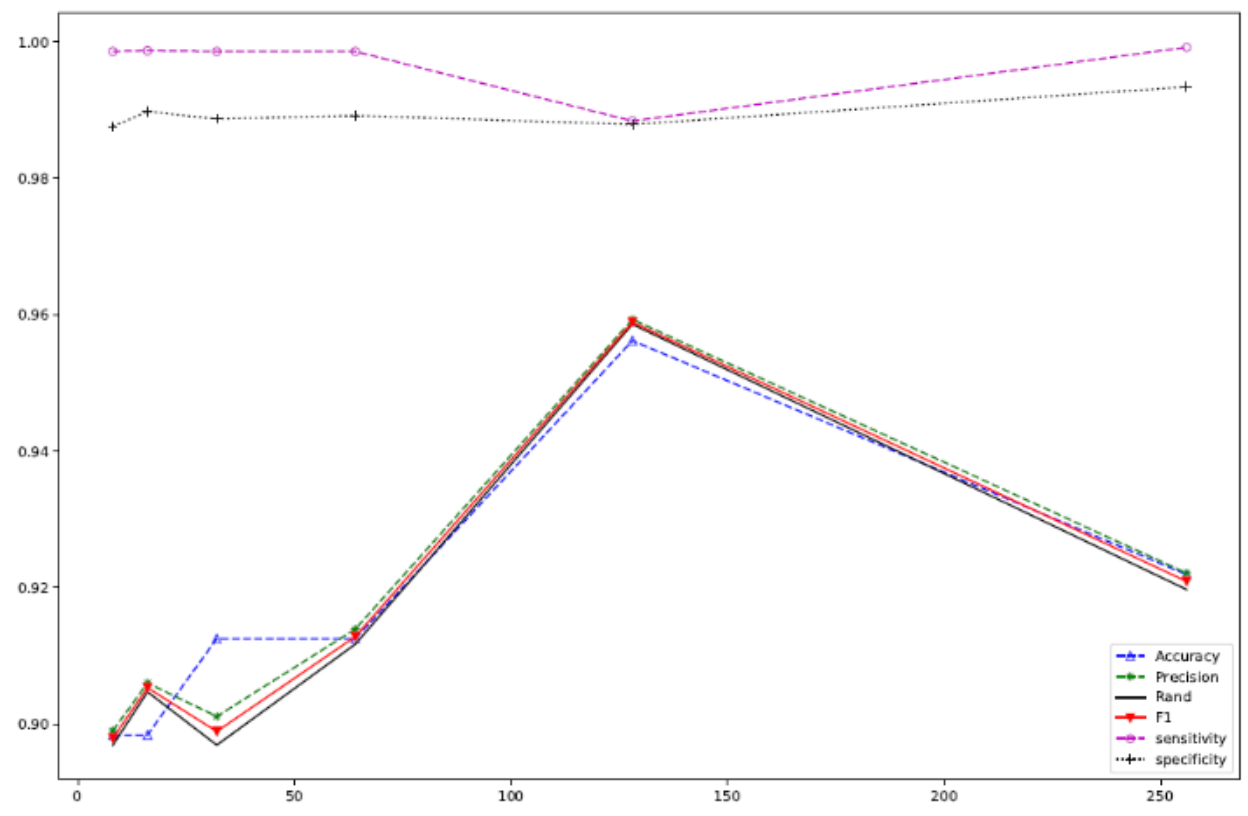

Figure 16: InceptionCapsule with self-attention

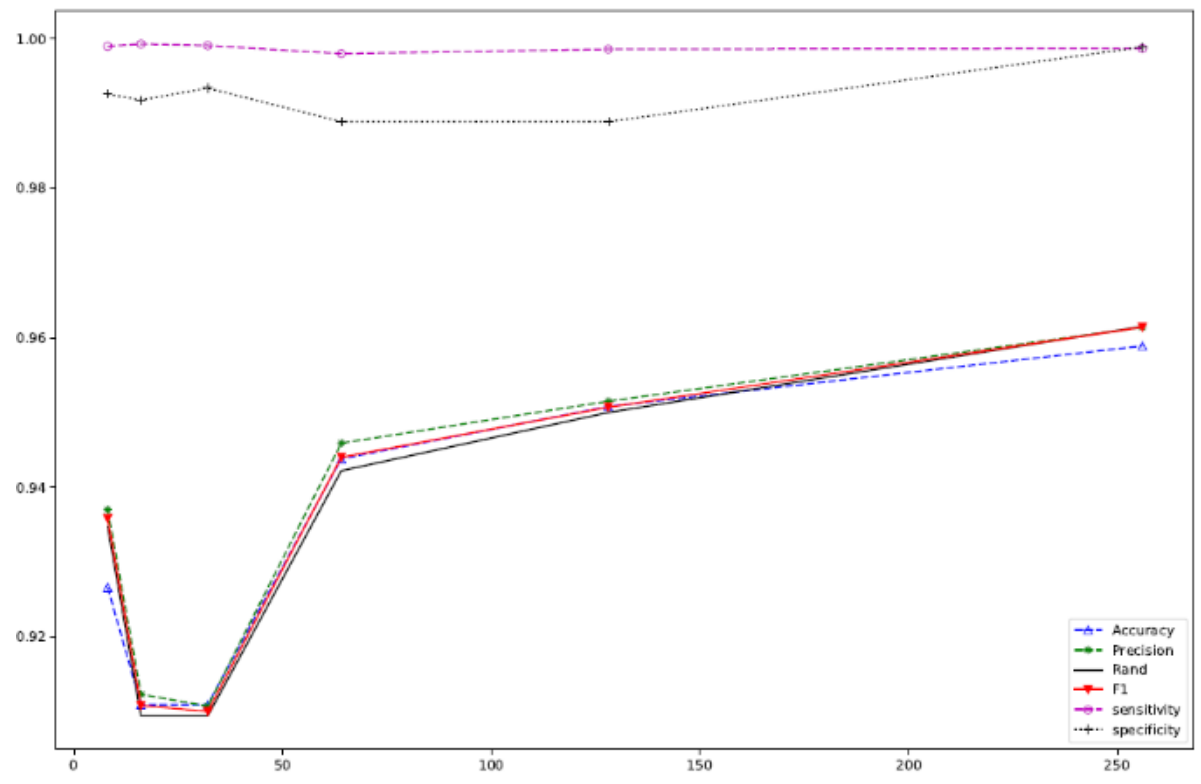

Figure 17:InceptionCapsule without self-attention

Figure 15: InceptionCapsule with self-attention and InceptionCapsule without self-attention batch size plots (8 class classification).

**4.3 BUSI Classification**

Table 4: Result obtained by proposed approach and other comparative models available in the literature.

| Model | Accuracy | Specificity | Precision | sensitivity | F1-score |
|---|---|---|---|---|---|
| **SD-CNN [56]** | 90.00 | 94.00 | - | 83.00 | - |
| **CNN-GTD [57]** | 86.50 | 88.02 | - | 85.10 | - |
| **GA-ANNs [58]** | 96.47 | 95.94 | - | 96.87 | - |
| **SeResNet1 [59]** | 90.00 | - | 91.00 | 90.00 | 91.00 |
| **Faster RCNN+CNNs [60]** | - | - | 87.60 | - | - |
| **CNN+LR [61]** | 96.87 | - | 86.93 | - | 82.64 |
| **ODET** | 97.84 | 98.63 | 93.28 | 93.96 | 93.16 |
| **Chowdary et al. (69)** | 88.08 | 86.13 | 85.62 | - | - |
| **Byra et al. [62]** | 92.33 | - | 81.00 | - | - |
| **SaTransformer[63]** | 93.34 | 88.32 | 89.51 | - | - |
| **InceptionCapsule with self-Attention** | 98.58 | 99.56 | 94.53 | 87.17 | 90.70 |
| **InceptionCapsule without self-Attention** | 98.88 | 99.53 | 95.34 | 91.20 | 93.74 |

The proposed approach was compared with 13 other common approaches in the literature. These approaches are SD-CNN [56], CNN-GTD [57], GA-ANNs[58], SeResNet18[59], Faster R-CNN+CNNs[60], and CNN+LR[61], Chowdary et al.[64], Byra et al.[62], Shi et al.[65], and SaTransformer[63]. Also, five criteria, Accuracy, specificity, precision, sensitivity, and F1-score were evaluated. Like other approaches in the literature, the proposed model used 0.80 data as training data and 0.20 data for model testing.

The SD-CNN approach was able to achieve Accuracy=90.00, Specificity=94.00, and Sensitivity=83.00. Other evaluation criteria are not available for this approach. The CNN-GTD approach was weaker than the SD-CNN and was able to record the values of Accuracy=86.50, Specificity=88.02, and Sensitivity=85.10. The rest of the metrics are not available for this approach. The GA-ANN approach recorded better results than the two reviewed approaches. This approach achieved Accuracy=96.47, Specificity=95.94, and Sensitivity=90.00. The SeResNet1 approach has worked similarly to SD-CNN in terms of accuracy and was able to reach Accuracy=90.00. This approach also reached Precision=91.00, sensitivity=90.00, and F1-score=91.00. In the Faster RCNN+CNNs approach, only the Precision metric was reported, which was equal to 97.60, and other metrics are not available for this approach. CNN+LR obtained a better accuracy value than other investigated approaches. This approach reached Accuracy=06.87, Precision=86.93, and F1-Score=82.64. The ODET approach obtained high metric values and has a high metric position compared to the examined approaches. This approach achieved Accuracy=97.84, Precision=93.28, Sensitivity=93.96, and F1-score=93.16.

The approach of Chowdary et al. [64] obtained weaker results than ODET; this approach was able to obtain Accuracy=88.08, Specificity=86.13, and Precision=85.62. The approach of Byra et al. [62] was able to reach the accuracy of 92.33 on the desired data set, and this approach also reached Precision=81.00. The SaTransformer approach[63] was also able to achieve 93.34 accuracy, which obtained weaker results than most comparative approaches. Also, this approach achieved Specificity=89.51, Precision=89.51, and sensitivity=89.51.

## 5. Conclusion and future works

In this work, a new framework for gastrointestinal classification based on Kvasir data and breast cancer classification based on BUSI data was introduced. The proposed framework was a capsule network with Self-attention mechanism that used transfer learning and Inception network to learn the initial vectors. The proposed approach used the Inception network to extract the initial vectors and solve the problem of creating the initial vectors. The ImageNet weights were chosen to avoid random weighting. Also, this approach used a capsule network to learn vectors better and an attention mechanism to select the best features. In this mechanism, Inception-ResNet was responsible for feature extraction and Capsule was responsible for learning and features. On the other hand, self-attention tries to select important features. Kvasir dataset was used to evaluate the proposed model, 20% testing, 10% validation and 70% training were considered. The proposed approach achieved comparable results with other approaches in the literature in two modes of 5-class classification and 8-class classification on Kvasir. The following steps are suggested as future works to improve the proposed approach:

- Using data augmentation, data generation, and re-sampling as a pre-processing step
- Remove the fully connected layer and use Bagging models
- Using other transfer learning structures such as efficient Net B0, efficient Net B1, ..., and efficient Net B7